\documentclass[conference,onecolumn]{IEEEtran}
\usepackage{amsmath}
\usepackage{amssymb}
\usepackage{amsfonts}
\usepackage{setspace}
\usepackage{comment}
\usepackage{cite}
\usepackage{graphicx}

\def\vector#1{\mbox{\boldmath $#1$}}

\date{}
\begin{document}

\title{
GPU accelerated Hybrid Tree Algorithm for Collision-less N-body Simulations
}

\author{
\authorblockN{Tsuyoshi Watanabe \& Naohito Nakasato\authorrefmark{1}}
\authorblockA{\authorrefmark{1}
Department of Computer Science and Engineering, University of Aizu\\
Aizu-Wakamatsu, Fukushima 965-0815, Japan\\
Email: nakasato@u-aizu.ac.jp}
}
 
\maketitle 

\section*{Abstract}
We propose a hybrid tree algorithm for reducing calculation and communication cost of 
collision-less N-body simulations.
The concept of our algorithm is that we split interaction force into two parts: 
hard-force from neighbor particles and soft-force from distant particles, 
and applying different time integration for the forces.
For hard-force calculation, we can efficiently reduce the calculation and communication cost 
of the parallel tree code because we only need data of neighbor particles for this part.
We implement the algorithm on GPU clusters to accelerate force calculation for both hard and soft force.
As the result of implementing the algorithm on GPU clusters, 
we were able to reduce the communication cost and the total execution time 
to 40\% and 80\% of that of a normal tree algorithm, respectively. 
In addition, the reduction factor relative the normal tree algorithm 
is smaller for large number of processes, 
and we expect that the execution time can be ultimately reduced down to about 70\% 
of the normal tree algorithm. 
  
\section{Introduction}
Gravitational N-body simulations deal with the motions of the many bodies (particles) 
interacting with other particles by gravitational force and are used for solving 
astronomical problems: formation of stars and galaxies.
There are basically two types of N-body systems: collisional and collision-less systems.
In a collisional system, a number of particles is relatively small and the orbit
of the particles is significantly deformed by force from the nearby particles.
In a collision-less system, a number of particles is large and the effect from near particles 
is relatively small. Also it does not necessary require highly accurate force calculation.
  
A most simple algorithm for calculating forces (or acceleration) between these bodies is 
a direct algorithm that calculates interactions of all $O(N^2)$ pair of particles.
However, we can reduce the calculation complexity by Barnes-Hut tree algorithm that 
approximates forces from many source particles as force from one source particle 
by tree structure for particles \cite{barnes1986}. 
The calculation complexity of the tree algorithm is $O(N{\rm log}N)$, 
but accuracy of force is worse at the expense of the approximation.
  
As the tree algorithm, many techniques for reducing calculation time of N-body simulation have been developed so far.
However, it is necessary to further speed-up the calculation for large-scale simulations. 
We usually speed-up N-body simulations by parallel computing using Message Passing Interface (MPI)
along with acceleration techniques such as Graphical Processing Units (GPU).
Nowadays, GPU is used for not only graphic processing but also general purpose processing. 
GPU enables us to accelerate N-body simulations 
by running the tree algorithm on it \cite{nakasato2011}.
  
As an approach for further reducing calculation cost of the tree algorithm, 
we can extend Particle-Particle Particle-Tree(PPPT) algorithm \cite{oshino2011}. 
PPPT algorithm is a hybrid of direct and tree algorithms for collisional simulation. 
In the method, we split gravitational force into short-rage and long-range force. 
The accurate direct algorithm is used for calculating short-range force
while we use the tree algorithm for calculating long-range force.
We apply different time integration methods for the two parts of the force.
Accordingly, we only adopt high accuracy methods for short-range force and 
can reduce the cost of unimportant (distant and weak) force calculation.
In this paper, we show a new algorithm based on PPPT scheme
for reducing calculation and communication cost of parallel N-body simulations.
We evaluated the performance of our algorithm on GPU clusters 
where each node of the cluster is equipped with GPUs.
  
\section{Hybrid Tree Algorithm} \label{sec:method}
In this section, we describe basic concepts for our hybrid tree algorithm.
  
\subsection{Gravitational N-body Calculation}
Motions of gravitational bodies follow the following equation of motion,
\begin{equation}
\frac{d^2 \vector{r}_i}{dt^2} = \sum_j^N F_{ij}.
\label{nbforce}
\end{equation}
Here, $F_{ij}$ is softened gravity force expressed as
\begin{equation}
F_{ij} = G m_{j} \frac{\vector{r}_j-\vector{r}_i}{(|\vector{r}_i-\vector{r}_j|^2 + \epsilon^2)^{3/2}},
\end{equation}
where $\vector{r}_i$ is a position of a "sink" particle that is forced from other particles, 
and $\vector{r}_j$ is a position of a "source" particle that exerts the force to other particles, 
$m_j$ is a mass of the sink particle, $G$ is a gravitational constant, 
and $\epsilon$ is the softening length to reduce non-realistic acceleration 
when $r_{ij} = |\vector{r}_i-\vector{r}_j| \sim 0$.
Simply, we calculate all pair interactions of particles for calculating right hand side of equation (\ref{nbforce}).
We call the simple algorithm for calculating force to a particle brute-force algorithm or direct algorithm. 
  
We solve the motion of particles by numerical integration of a position of each particle with the calculated force. 
Actually, the integration is performed by updating velocities followed by updating positions of particles.
This integration scheme is called the leap-frog method.
The leap-frog scheme is a second-order symplectic integrator.
A velocity and a position of a particle are updated as follow,
\begin{eqnarray}
  	\vector{v}_{t+1/2}= \vector{v}_{t-1/2}+\Delta t\vector{a}(\vector{r}_{t})\\
  	\vector{r}_{t+1}=\vector{r}_{t}+\Delta t\vector{v}_{t+1/2}
\end{eqnarray}
where $\vector{v}_{t}$ is velocity of a particles at time $t$, 
$\vector{r}_{t}$ is position of a particle at time $t$, $\Delta t$ is a time-step for integration. 
In the following, we call the velocity update as "kick" and the position update as "drift". 
  
\subsection{Tree Algorithm} \label{sec:tree}
The tree algorithm is a technique for reducing the cost of force calculation 
for large-scale simulations \cite{barnes1986}. 
The concept of tree algorithm is that we approximate force from many distant source particles into force 
from one source particle as the center of mass of the particles.
Force calculation by tree algorithm is performed as follow: 
constructing tree; calculating center of mass of tree nodes and criterion for depth of tree traversal; and traversing tree and calculating force.
  
For constructing tree structure, we divide three dimensional space into eight equal size cells recursively 
from root cell that contains all particle in the system. 
The division is recursively continued while the cell has many particles than 
a critical number of particles $n_{ \rm crit}$. 
As the result, the particles are placed on leaves of the tree.
  
Next, to approximate distant particles, we calculate center of mass of cell for each cell.
  
Then, we calculate multi-pole acceptance criterion (MAC) of each tree node as the criterion for tree traversal. 
MAC determines whether we further traverse leaf cells of the cell or calculate force from the cell. 
We use Absolute MAC \cite{warren1991},
\begin{eqnarray}
  r_b = \frac{b_{ \rm max}}{2} + \sqrt{\frac{b_{ \rm max}^2}{4}+\sqrt{\frac{3B_2}{a_{ \rm err}}}}\\
  B_2 = \sum_{i}^{N_{ \rm cell}} m_i | {r}_{ \rm CM} - {r_i} |^2
\end{eqnarray}
where $b_{ \rm max}$ is the maximum distance between the center of mass and particles in the cell, 
$r_{ \rm CM}$ is a position of a center of mass of source cell, 
and $N_{ \rm cell }$ is the number of particles in the cell.
$a_{ \rm err}$ is a numerical parameter specified by user to control
the accuracy of force calculation.
   
Finally, we traverse the tree for calculating force. 
We start from root cell. 
If $r_b > r_{ij}$, where $r_{ij}$ is a distance between sink particle and 
center of mass of source cell, we further visit to leaf cells
to traverse in more detail, else we add the force from the cell and go to next node. 
After tree traversal, we get the force of a sink particle by summing forces from source cells and particles.

\subsection{Hybrid Tree Algorithm}
Here, we explain an algorithm we proposed based on PPPT algorithm for collision-less systems. 
Our algorithm is the similar to PPPT algorithm: splitting force into hard-force from near particles 
and soft-force from distant particles but we adopt a different numerical method for the hard-force part.
In the original PPPT algorithm, the direct algorithm is used for high accuracy calculation of hard-force.
The high accuracy is not necessary for collision-less simulation, 
thus we can design our algorithm for hard-force having lower accuracy than original PPPT algorithm.
Another difference of our algorithm is that we try to speed-up the calculation 
by reducing communication cost in parallel computing.
  
The force is divided by using a kernel function $K(r_{ij})$ as follow.
\begin{eqnarray}
   F_{ij}=F_{ij,\rm Hard}+F_{ij,\rm Soft}\\
   F_{ij,\rm Hard} = F_{ij} K(r_{ij})\\
   F_{ij,\rm Soft} = F_{ij} (1-K(r_{ij}))
\end{eqnarray}
where $F_{ij,\rm Hard}$ is the hard-force, and $F_{ij,\rm Soft}$ is the soft-force.
For a kernel function, we use the DLL function (adopted in \cite{oshino2011}) written as follows,
\begin{equation}
  		   K(r_{ij})=\left\lbrace \begin{array}{l l}
  		   1& (Y\geq 1),\\
  		   10Y^6-15Y^8+6Y^{10} & (0<Y< 1),\\
  		   0& (Y\leq 0),\\
  		   \end{array}
  		   \right.
\end{equation}
where $Y=\frac{R_2-r_{ij}}{R_2-R_1}$, $R_1$ and $R_2$ are constants specified 
by user determining the size of transition zone between hard and soft forces. 
  		    
\begin{figure}[htbp]
\centering
\includegraphics[width=0.9\linewidth]{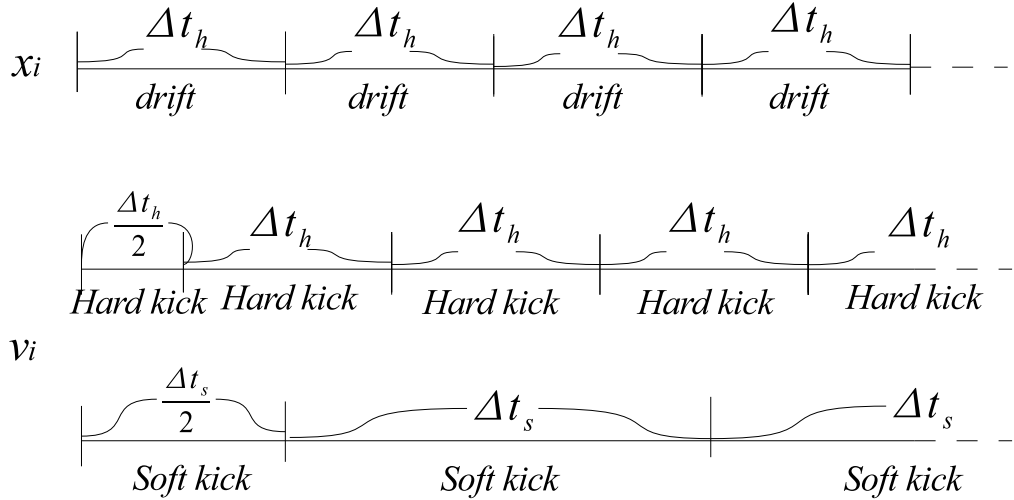}
\caption{Illustration of our integration scheme with $n=2$.}
\label{fig:ht}
\end{figure}

We use the tree algorithm and the leap-frog for both of soft and hard forces
but the time-step for soft ($\Delta t_s$) and hard ($\Delta t_h$) are
different.
We make the relation between $\Delta t_s$ and $\Delta t_h$
as $\Delta t_s = n \Delta t_h$, where $n \in \mathbb{N}$. 
Illustration of our integration is shown in Figure \ref{fig:ht}.
We call the step calculating both of soft and hard force "soft-step" and 
the step of calculating only hard-force "hard-step"; 
we need one soft-step and $n-1$ hard-steps to calculate time evolution for $\Delta t_{\rm s}$.
Since the soft-step theoretically require the position of all particles but the hard-step only relies
on the position of near particles, we expect that calculation and communication cost of the proposed hybrid
tree algorithm is lower than the normal tree algorithm.
The time integration error of our algorithm expected to be slightly larger than the normal tree algorithm 
with time-step $t_{ \rm h}$ due to reduction of long-range force calculation. 
However, we can control the error by choosing appropriate parameters 
for $\Delta t_s$, $\Delta t_h$, $n$, $R_1$, and $R_2$.		

\section{Parallel Implementation}\label{sec:imple}
In this section, we present how we use make our parallel using OpenMP and MPI
along with explanation for GPU computing.

\subsection{Parallel Tree Algorithm with GPU}
In this section, we show the procedure and data structure for our tree algorithm. 
Our method is based on \cite{nakasato2011} as constructing tree by CPU and traversing tree by GPUs.
  
\subsubsection{Tree construction on CPU}
First, we construct tree structure of particles. 
We make cell-nodes above the particle-nodes and connect the nodes with pointers.
In the method, each node has "more" pointer to the first leaf cell
and "next" pointer to the next cell/particle to traverse skipping over leafs.
  
For tree construction, we first calculate the region and size of the root cell 
that include all particles.
Then, we calculate keys of the particles.
We use the Morton key as the key that is following order of Morton curve (or Z-curve), a space filling curve.
The advantage of Moron key is that it encodes hierarchical information of position of particles.
The key calculation is able to be executed individually for particles, 
thus we use OpenMP for parallelizing the calculation.
Second, we sort the keys. 
We use the extension of C++ standard library std::sort for sorting in parallel with OpenMP.
We also sort the data of the positions of the particles to preserve locality of the particles.
Third, we divide the array into eight sub-arrays by the three most significant bits of the keys, 
then we set a cell node with "more" pointer and "next" pointer and make child cells with next pointers. 
Then we recursively repeat the procedure for every three bits of key while the array has $n_{ \rm crit}$ 
or more particles.
If the array has particles fewer than $n_{\rm crit}$, we treat each particle as leaf node.
Finally, we need to calculate center of mass and MAC of each cell. 
Both of them are calculated from position and mass of particles contained in the cell. 
Thus, we calculate center of mass and MAC by traversal of the part of tree, 
and the calculation is individual for each cell, and we parallelize the calculations by OpenMP.
  
\subsubsection{Tree Traversal on GPUs}
We use GPUs for tree traversal and force calculation. 
This part is implemented in OpenCL, the framework for parallel computing. 
First, CPU sends the data of tree to GPU. 
Then, we run the kernel code for traversing tree.
The kernel code traverse the tree by indexing ``more'' and ``next'' pointers.
The tree traversal is individually executed for each sink particle. 
Thus, all threads are run by GPUs in parallel.
In addition, for reducing a number of tree traversal, multiple sink particles traverse in same thread. 
The number of particles traversing in same thread is $n_{ \rm vec}$. 
We typically set $n_{ \rm vec} = 4$ as efficient number of particles for GPU. 
As a distance for determining whether traverse the children or not, 
we use minimum distance between a source cell $j$ and $n_{ \rm vec}$ sink particles $i$; 
we traverse the leaf if
\begin{equation}
  r_{\rm b} > {\rm min}(r_{ij})
\end{equation}
for $0\le i < n_{\rm vec}$. If particles are unsorted, and $n_{ \rm vec}$ particles are distant each other, 
we need to traverse unnecessary nodes because ${\rm min}(r_{ ij})$ may be small for distant cell. 
Thus, for reducing unnecessary traversal, we should sort the particle data so that 
we retain data locality of positions of particles.
  
\subsection{MPI Parallelization}
Our method of parallelization is that we assign each MPI process own region that contains 
subset of particles, 
and each MPI process calculates force by own particles and particles received from other processes.
We have already presented the parallelization on each process attached GPU.
Here, we show the implementation of parallel computation and communication 
of our algorithm on GPU clusters. 
Our procedure for parallel N-body simulations is as follow,
\begin{enumerate}
   \setlength{\parskip}{0cm}
    \setlength{\itemsep}{0cm} 
  \item Domain decomposition
  \item Constructing local tree
  \item Calculating force from local tree
  \item Communicating tree from remote processes
  \item Calculating force from remote tree
  \item Updating positions and velocities of local particles
\end{enumerate}
  
\subsubsection{Domain Decomposition}\label{sec:dd}
First, we need to distribute particle data to each process. 
To simplify communication for hard-force calculation, 
a shape of a region of a process should be a cuboid. 
As the method of cuboid domain decomposition, we use the method introduced in \cite{makino2004}. 
With the method, we decompose whole region into $P=P_{x} \times P_{y} \times P_{z}$ regions, 
where $P_{x}, P_{y},$ and$ P_{z}$ are the number of division in $x, y,$ and $z$ direction. 
The decomposition is implemented as exchanging of particles between neighbor processes 
given pre-determined boundaries between regions.
  
To determine the boundaries, we use the sampling domain decomposition method used in \cite{ishiyama2009}.
In the method, we gather sample particles to a main process, 
then the main process tries to balance the boundaries such that 
each process has approximately same number of sample particles.

\begin{figure}[htbp]
\centering
\includegraphics[width=0.9\linewidth]{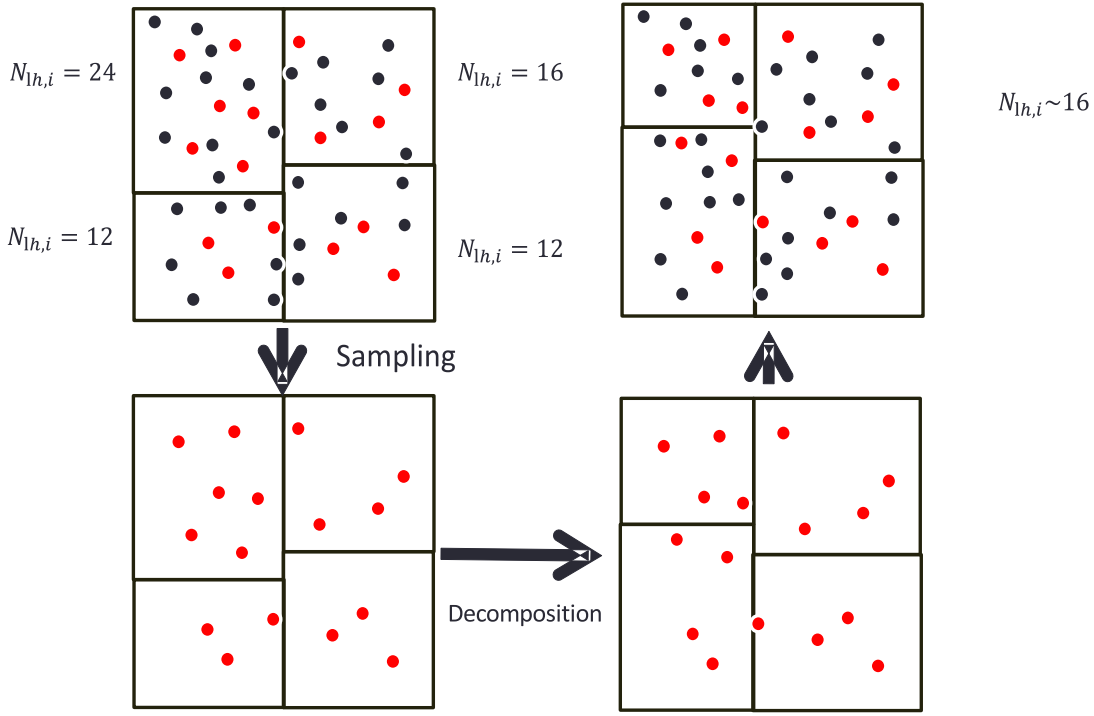}
\caption{Illustration of our domain decomposition 
where $N_{\rm lh, \it i}=N_{ \rm local,\it i} +N_{ \rm hard,\it i}$.}
\label{fig:dd}
\end{figure}
  
Illustration of sampling domain decomposition is shown in Figure \ref{fig:dd}.
Here, the number of sample particles $N_{ \rm samp}$
is defined for balancing the sum of the number of local particles
that are assigned to each process $N_{ \rm local,\it i}$ and 
the number of particles received from other processes for 
hard-force calculation $N_{ \rm hard,\it i}$.
$N_{ \rm samp}$ for process $i$ is determined as follow;
\begin{equation}
  N_{ \rm samp,\it i} = N R_{ \rm samp} f_{\rm samp,\it i},
\end{equation}
where $N$ is the total number of particles, $R_{ \rm samp}$ is sampling rate constant, 
and $f_{\rm samp,\it i}$ is a correction factor for balancing. 
We typically set $R_{ \rm samp} = 2^{-8}$ in the present work.
$f_{ \rm samp,\it i}$ is the measure for load balancing defined as
\begin{equation}
  f_{ \rm samp,\it i}=\frac {N_{ \rm local,\it i} +N_{ \rm hard,\it i}}{N}.
\end{equation}
Our intention is that we make the calculation cost for hard-force equal on all processes
because the calculation is the majority of running time in our case.
After the main process determines the boundaries, it broadcasts the result to all other processes,
and each process exchanges necessary particle data between other processes.
To reduce the cost of domain decomposition, we execute it only for every soft-step; 
at hard-steps, a process has the same particles as the previous soft-step.
  
\subsubsection{Communication of Particles}
After construction of tree structure of local particles, 
we need to communicate the particles of other processes to calculate the force from the particles. 
In our method, we need different set of particles for hard and soft force, respectively.
  
For communicating the soft particles, we need data of all particles, 
but distant particles are able to be approximated as the center of mass of the cells.
Locally Essential Tree (LET) is the method for communicating only necessary part of 
tree for the processes \cite{warren1992}. 
For determining the cells to send, we traverses the local tree with MAC,
  \begin{equation}
  r_{i\rm b_{\it j}}<r_{\rm b} \label{eqn:mac},
  \end{equation}
where $r_{i\rm b_{\it j}}$ is the distance between center of mass of cell $i$ 
and boundary of process $j$, and $r_{\rm b}$ is the MAC calculated by method in Section \ref{sec:tree}.
We only send the position and the mass of center of mass for a cell. 

\begin{figure}[htbp]
\centering
\includegraphics[width=0.9\linewidth]{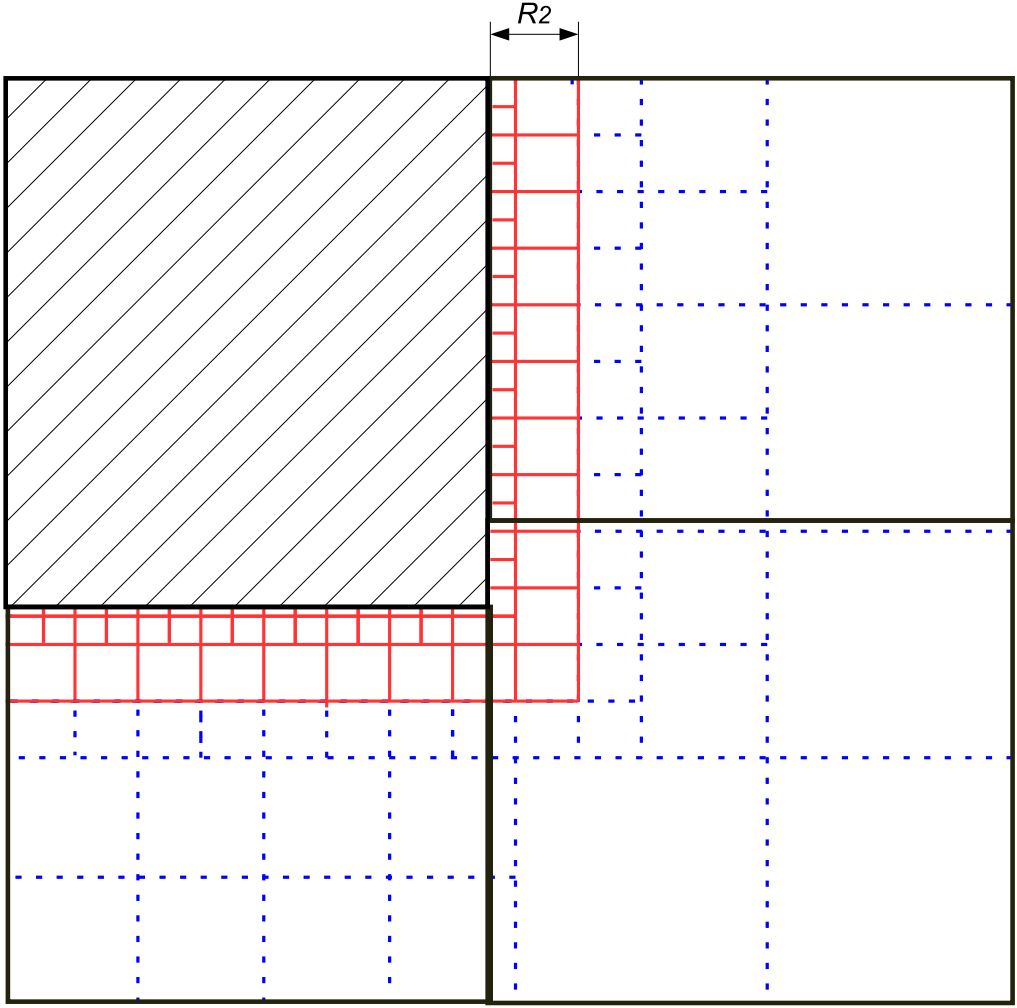}
\caption{Illustration of communication of cells with LET.
We only need red cells for hard-force and additionally use blue dashed
cells for soft-force.}
\label{fig:comm2}
\end{figure}

Both the calculation cost for determining cells to send and 
the cost for communicating cells are $O(P{\rm log}{N_{\rm local}})$.
As the result of communication, a process get the cells that we need to calculate 
the force in the process as shown in Figure \ref{fig:comm2}. 
Here, we show the cells that upper-left process needs to receive from other processes.
We use MPI asynchronous send and receive functions to exchange data.
After the communication, the process concatenates the arrays of 
own particles and the cells received by neighbor processes and constructs a LET.
Then, we traverse the tree and calculate force 
from the remote particles to local particles. 

For hard-force, we also use LET scheme, but a process only need cells around the boundary that is at
the distance less than $R_2$ as showing in the red cells in Figure \ref{fig:comm2}.
To determine the cells to send, we traverse the local tree. 
In addition to MAC in Equation (\ref{eqn:mac}), 
the condition to determine whether traverse the leaf for searching the cells or not 
is applied as follow
\begin{equation}
  r_{ i\rm b_{\it j}}<R_2 + b_{ \rm max}.
\end{equation}
After the tree traversal, we obtain the cells in the process 
that the distance to boundary of process $j$ is smaller than $R_2$. 
The communication cost for hard-force is smaller than for soft-force.
Especially, the cost is significantly reduced in large number of processes 
because of reduction of volume that a process needs to consider.
  
\subsubsection{Overlap of Force Calculation and Other Calculations} \label{sec:ol}
We execute the force calculation on GPUs and other calculations on host CPU. 
Thus, we overlap both calculation with communications.
While traversing the tree in GPUs, CPU communicate particles and construct a LET using received particles. 
While GPUs run kernels, we need to retain a thread for management of the GPU. 
The thread is generated by using pthread API, an programming interface the interface for thread programming.
As the result of the overlap, the total calculation time of the overlapped processes is constrained 
by the maximum calculation time of the CPU threads and the GPU. 
However, we need a CPU thread for organizing queuing jobs to OpenCL device such as GPU. 
Thus, performance of calculations that use OpenMP may be decreased. 

\section{Results}\label{sec:result}
In this section, we present the performance evaluation of N-body simulations 
with our hybrid tree method. 
  
\subsection{Settings of Simulation}\label{sec:setting}
For the test of our algorithm, we use Plummer model \cite{plummer1911}. 
The Plummer model is a typical spherical model of N-body simulations. 
We set $\epsilon=0.1 N^{-0.26}$, and $\Delta t_{\rm h} = \frac{\epsilon}{\sigma_v}$, where $\sigma_v$ 
is mean velocity of the system in the case of our simulation $\sigma_v \sim 0.65$, 
as being in range of optimal parameters for the model shown in \cite{rodionov2005}. 

We have the following numerical parameters that control the balance between
the execution time and accuracy of the simulations: $a_{err}$, $n$, $R_1$, and $R_2$. 
In the present work, we typically set $a_{\rm err}=2^{-8}$. 
$n$, $R_1$, and $R_2$ should be adjusted for maintaining sufficient accuracy of error in total energy 
of the system that is the sum of kinetic and potential energy after simulation. 
By the result of test simulations for our model, we choose an optimal parameters as
$n=4$, $R_1=1.03N^{-\frac{1}{3}}$, and we sent $R_2=2R_1$ in the present work.

Development and computations for the present work have been carried out 
under the ``Interdisciplinary Computational Science Program'' 
in Center for Computational Sciences, University of Tsukuba.
A node of HA-PACS has two Intel E5-2670 (8 cores) CPUs with four NVIDIA Tesla M2090 GPUs.
Actually, we assign four MPI processes per node of HA-PACS such that
one MPI process is exclusively assigned one GPU board. 
  
\subsection{Reduction of Calculation and Communication} \label{sec:red}
Here, we compare the calculation cost of kernel for calculating hard-force by our algorithm 
and communication cost with the normal tree algorithm with LET. 
  
In hard-force calculation, we can cut-off the tree traversal for distant cells. 
Thus, the cost for hard calculation is reduced if we set small $R_1$ and $R_2$. 
As the result of our test simulations, for $N$ between 256k and 4096k, 
the time for calculating hard force at optimal $R_1=1.03 N^{-\frac{1}{3}}$ 
is about 40\% of the time for calculating force with normal tree algorithm.

\begin{figure}[htbp]
\centering
\includegraphics[width=0.9\linewidth]{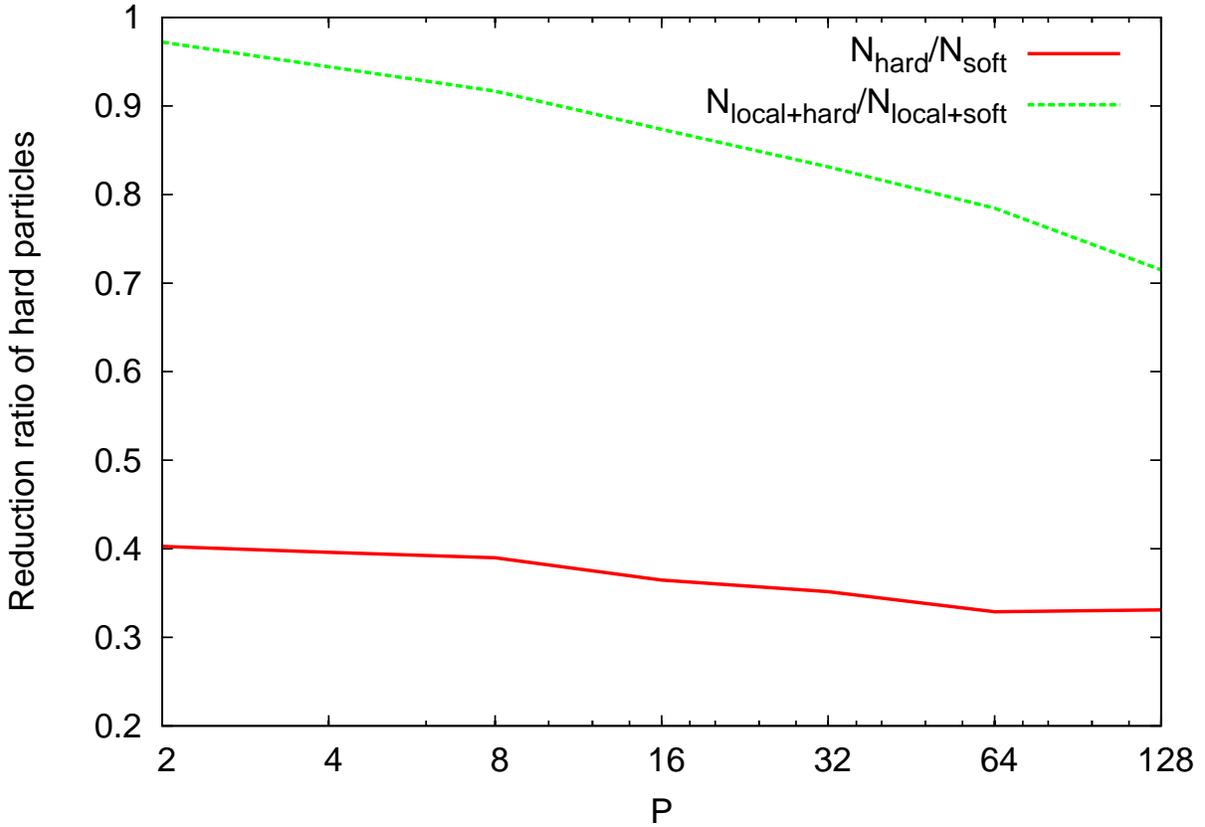}
\caption{$N_{\rm hard}/N_{\rm soft}$ and $N_{\rm local+hard}/N_{\rm local+soft}$ as a function of $P$.}
\label{fig:nphs}
\end{figure}

Next, we analyze the cost for communicating of our algorithm with the normal tree algorithm.
For the test, we set $N=8192 \rm k$ (${\rm k}=2^{10}$).
In Figure \ref{fig:nphs}, the solid red line shows the ratio between the average number of 
hard particles $N_{\rm hard}$ and soft particles $N_{\rm soft}$ as a function of $P$, the number of processes.
This ratio is an indicator of the reduction of cost for GPU computing and is roughly constant at 40\%.
The dotted green line shows the ratio between the average number of local plus hard particles $N_{\rm local+hard}$
and local and soft particles $N_{\rm local+soft}$.
Since the communication cost for our algorithm and the normal tree
are proportional to $N_{\rm local+hard}$ and $N_{\rm local+soft}$, respectively, 
we see that our algorithm works better in large $P$ due to the reduction of communication.
For larger $P$, we have smaller the ratio as $\sim$ 70 \% at $P$ = 128.
  
\subsection{Scalability of Hybrid Tree Simulation}\label{sec:scal}
We evaluate the scalability of our simulation with GPU clusters.
It is not easy to reduce the execution time by number of processes linearly, e.g. good strong scaling, 
even if our hybrid tree algorithm can reduce the communication cost by 
reducing the volume of interest for communication.
For the test, we run a series of simulations with $N =$ 1M (M = $2^{20}$) to 64M 
on up to $P=128$ using 32 nodes of HA-PACS.

\begin{figure}[htbp]
  \centering
  \includegraphics[width=0.9\linewidth]{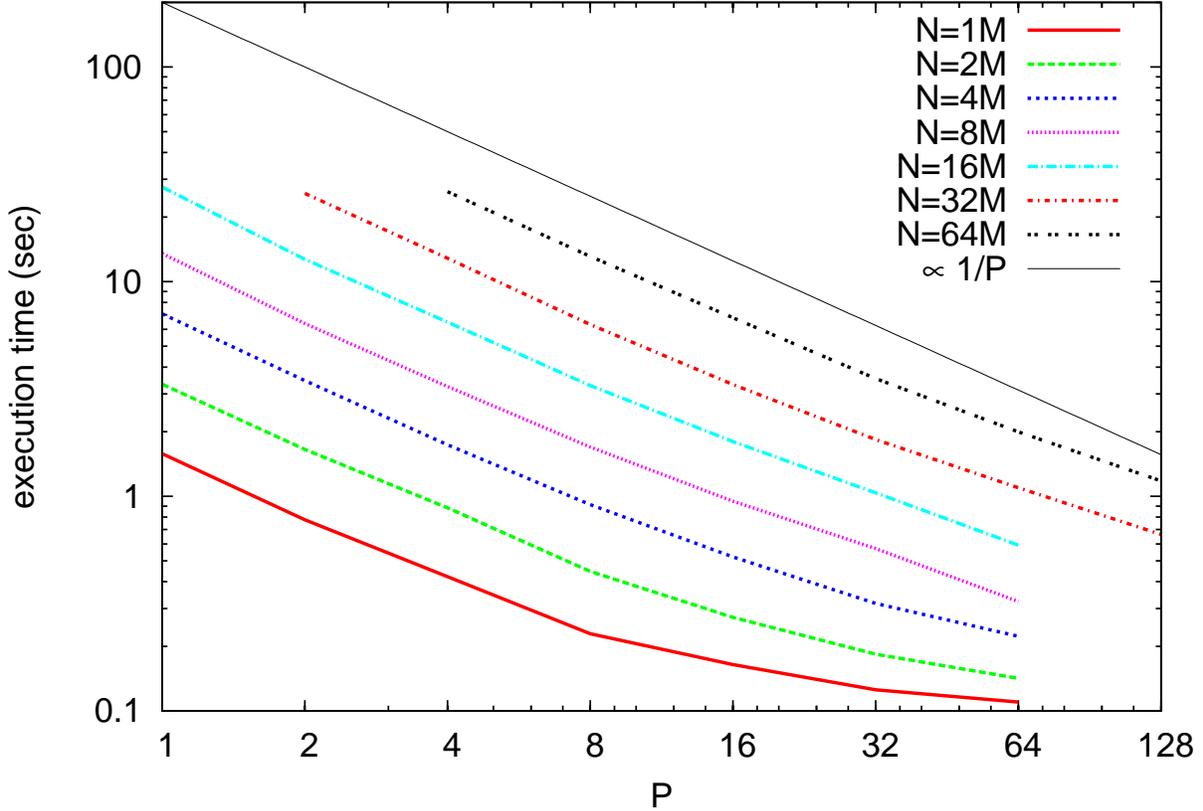}
  \caption{Strong scaling of our hybrid tree algorithm on HA-PACS.}
  \label{fig:scaltime}
\end{figure}
  
Figure \ref{fig:scaltime} shows the strong scaling result of our simulation; 
capability of the speed-up with many processes for fixed total number of particles. 
Here, we plot the average execution time for simulating $\Delta t_{\rm s}$ time evolution 
as a function of $P$.
The time evolution of $\Delta t_{\rm s}$ is completed with one soft-step and three hard-steps
in the present work (we set $n = 4$).
We omit some cases of the simulations that were not able to run due to the limitation of GPU memory in the figure.
The execution time is reduced in $O(1/P)$ for small $P$, but the time hardly reduces at large $P$ and small $N$. 
For $N=1\rm M$, the reduction of the execution time stops at $P=64$. 
For $N=64\rm M$, the execution time is reduced approximately linearly, 
and the time at $P=128$ is 59\% of the time at $P=64$. 
As the result, the execution time is sufficiently reduced when $N/P \gtrsim 1 {\rm M}$; 
the calculation time with $2P$ processes is typically less than 60\% of that with $P$ processes.
  
\begin{figure}[htbp]
\centering
\includegraphics[width=0.9\linewidth]{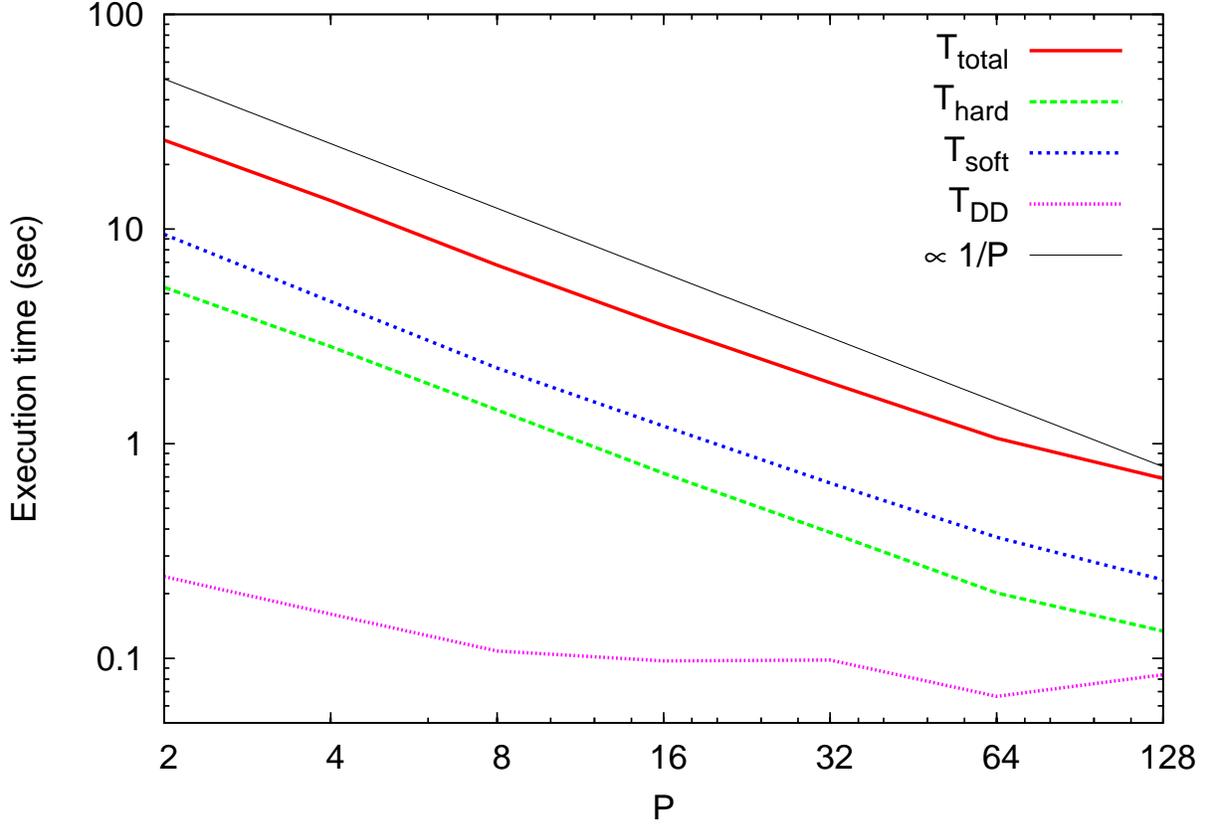}
\caption{Detailed breakdown of the execution time.}
\label{fig:dettimet}
\end{figure}

In Figure \ref{fig:dettimet}, we present the detailed breakdown of the execution time 
for hard-step, soft-step, and domain decomposition for the simulation with $N=32\rm M$.
Here, $T_{\rm total}$, $T_{\rm hard}$, $T_{\rm soft}$, and $T_{\rm DD}$ are 
the execution times of total for simulating $\Delta t_{\rm s}$, 
one hard-step, one soft-step, and domain decomposition, respectively. 
The relation between those timing is expressed as
$T_{\rm total} = 3T_{\rm hard} + T_{\rm soft} + T_{\rm DD}$. 
$T_{\rm hard}$ is reduced to about 60\% of $T_{\rm soft}$ for any $P$. 
$T_{\rm hard}$ and $T_{\rm soft}$ are reduced as increasing of $P$. 
In $T_{\rm hard}$ for $P=128$, time for force calculation, tree construction, and communicating cells 
are 26\%, 37\%, and 54\%, respectively. 
The sum of percentages of time is larger than 100\% 
because kernel execution on GPU and other processes on CPU are overlapped.
$T_{\rm DD}$ is not reduced and be around 0.1 seconds for this case.
The reason is that communication and calculation cost for domain decomposition depends 
on not $P$ but $N$ as shown in Section \ref{sec:dd}. 
  
Since the core of our tree code is written in OpenCL API, 
we can use not only GPUs but CPU threads to compute 
the tree travarsal kernels for hard and soft force.
For $N = 32$M runs, the total execution time with $P = 4, 16, 128$ is 113, 30.6, and 5.51
seconds, respectively while the runs with GPUs took 13.6, 3.55, and 0.688 seconds.
The speed-up factor due using GPUs with $P = 4, 16, 128$ is 8.3, 8.6, and 8.0, respectively
where we compare the time for all computation and comunication.
To be more specific only on computation, we found the speed-up
factor of the execution of OpenCL kernels is 11 - 16 times faster than the runs with CPU threads.
Our hybrid tree algorithm can take huge advantage of the acclearation with the GPU technology.
    
\subsection{Comparison to Normal Tree Algorithm}
Here, we compare the result of the execution time of our algorithm to the normal tree algorithm with LET 
that does not split force into two parts.
To achieve approximately same total energy error between two algorithms,
we set $\Delta t_{\rm h}=\Delta t_{\rm n}$, where $\Delta t_{\rm n}$ is the time-step of the normal tree. 
In addition, domain decomposition in the normal tree is executed every four steps to fairly compare the execution time.
  
\begin{figure}[htbp]
\centering
\includegraphics[width=0.9\linewidth]{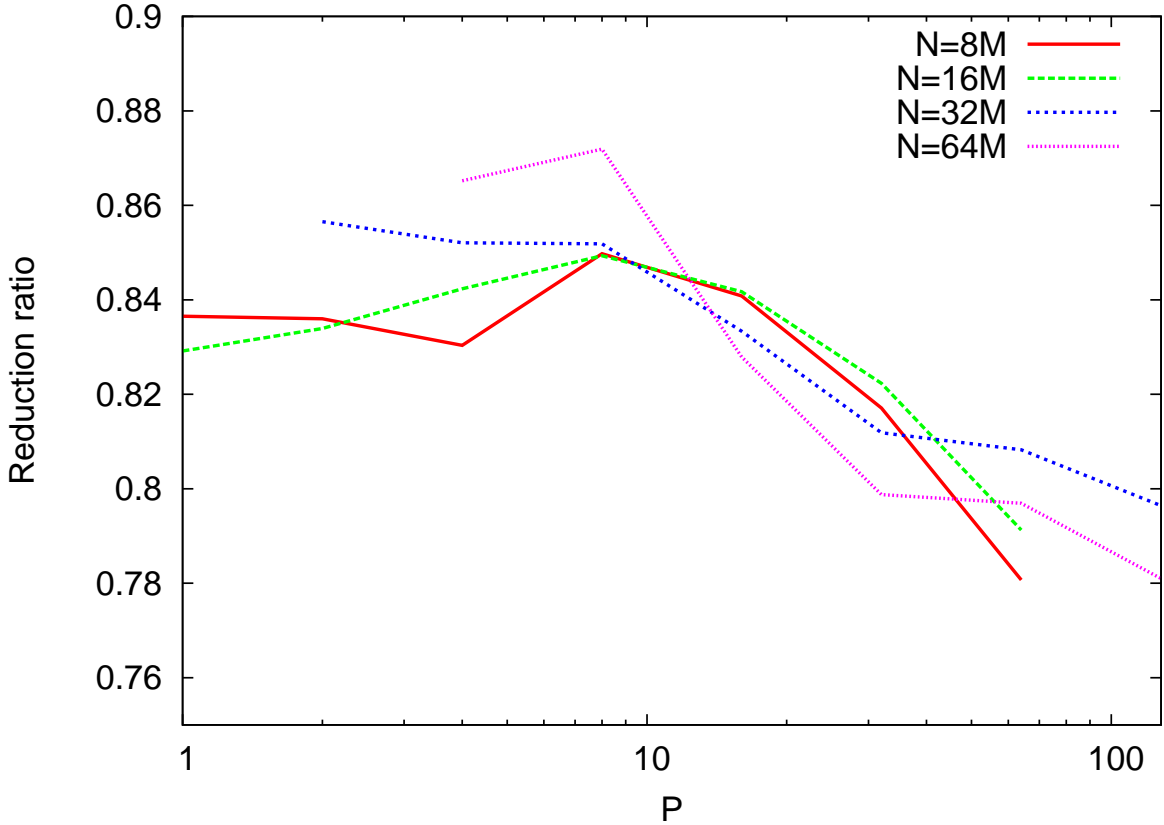}
\caption{Reduction of the execution time of our algorithm.}
\label{fig:ration}
\end{figure}

Figure \ref{fig:ration} shows the reduction of the execution time of the hybrid tree algorithm 
versus the normal tree algorithm $R_{\rm hybrid}=T_{\rm hybrid}/T_{\rm normal}$, 
where $T_{\rm hybrid}$ and $T_{\rm normal}$ is the execution time of the hybrid tree 
and the normal tree for same simulation time $\Delta t_{\rm s} = 4\Delta t_{\rm n}$.
We can reduce the time to about 80\% - 90\% of that of the normal tree. 
Especially, for $P>8$, $R_{\rm hybrid}$ is even smaller as $P$ is larger. 
This means that our hybrid tree algorithm has the advantage for large-scale simulations.
  
The theoretical reduction of the hybrid tree $R_{\rm t,hybrid}$ is estimated as
\begin{equation}
R_{\rm t,hybrid}=\frac{(n-1) T_{\rm hard} + T_{\rm soft}}{n T_{\rm tree}},
\end{equation}
where $T_{\rm tree}$ is the execution time of the normal tree algorithm.
According to the results in section \ref{sec:red} and \ref{sec:scal}, 
it is expected that the hybrid tree algorithm can reduce the cost for hard-force
to about 60 \% of soft-force for large $P$.
Thus, assuming that $T_{\rm soft}=T_{\rm tree}$, $T_{\rm hard} = 0.6 T_{\rm tree}$, and $n=4$, 
then $R_{\rm t,hybrid}=0.7$; we can ultimately speed-up the calculation 
with hybrid tree to 70\% of the normal tree for large $P$ 
except for time for the domain decomposition.

\subsection{Comparison to Other Work}
Ogiya et al. implemented parallel tree N-body code on HA-PACS \cite{ogiya2013}. 
Their GPU code use the same algorithm \cite{nakasato2011} also used in the present work.
However, the detailed implementation details of their tree traversal kernels
and domain decomposition are different.
In \cite{ogiya2013}, they presented a model of CDM (Cold Dark Matter), 
and they claimed it was hard to keep load balance when $P>8$.
For $N=32\rm M$ and $P=8$, the execution time for four time-steps in \cite{ogiya2013} $T_{\rm o}=$ 8.2 seconds 
and the execution time for three hard steps and one soft step in our work $T_{\rm hybrid}=$ 6.8 seconds 
and $\frac{T_{\rm hybrid}}{T_{\rm o}}$=0.83. 
For $P=64$, $T_{\rm o}=$ 3.0 seconds and $T_{\rm hybrid}=$ 1.1 seconds 
and $\frac{T_{\rm hybrid}}{T_{\rm o}}$=0.37. 
Although the implementation and a simulation model are different to \cite{ogiya2013}, 
our algorithm can efficiently reduce execution time for scalable computation.
  
\section{Conclusion}
In this work, we developed a new algorithm for N-body simulations named hybrid tree algorithm, 
the algorithm for accelerating collision-less N-body simulation by splitting
the force from other particles into short-range and long-range forces. 
The proposed hybrid tree algorithm is effective to reduce the calculation 
cost and communication cost for simulations.
We have implemented the algorithm on GPU clusters up to 128 processes, 
and we showed that the hybrid tree algorithm can reduce the execution time up 
to 80\% of the normal tree algorithm. 
As future work, we should investigate the scalability and speed-up of 
our algorithm with more scalable computing systems. 
In addition, we will investigate whether our algorithm is efficient 
for other systems and other parameters because we have simulated 
the algorithm with only limited combinations of parameters and only on Plummer model.

\end{document}